%% file: paper.tex
\documentclass[conference,10pt]{IEEEtran}
\IEEEoverridecommandlockouts
\usepackage[utf8]{inputenc}
\usepackage[T1]{fontenc}
\usepackage{graphicx}
\usepackage{amsmath,amssymb}
\usepackage{booktabs}
\usepackage{enumitem}
\usepackage{hyperref}
\usepackage[table]{xcolor}
\definecolor{autogategreen}{HTML}{E1F1EC}
\definecolor{basegray}{HTML}{F5F5F5}
\usepackage{listings}
\usepackage{cite}
\usepackage{microtype}
\usepackage{tikz}
\usetikzlibrary{arrows.meta,positioning,shapes.geometric,fit,calc,backgrounds}
\usepackage{stfloats}
\usepackage{balance}
\usepackage{multirow}
\usepackage{makecell}
\usepackage{tabularx}
\usepackage{algorithm}
\usepackage{algorithmic}
\usepackage{pifont}
\usepackage{xspace}
\usepackage{wrapfig}
\usepackage{colortbl}
\usepackage{caption}
\newcommand{\cmark}{\textcolor{green!60!black}{\ding{51}}}
\newcommand{\xmark}{\textcolor{red}{\ding{55}}}

\newcommand{\autogate}{\textsc{AutoGate}\xspace}

\makeatletter

\long\def\@makecaption#1#2{%
\ifx\@captype\@IEEEtablestring%
\footnotesize\bgroup\par\raggedright\@IEEEtabletopskipstrut{\normalfont\footnotesize #1.\nobreakspace #2}\par\addvspace{0.5\baselineskip}\egroup%
\@IEEEtablecaptionsepspace%
\else%
\@IEEEfigurecaptionsepspace%
\setbox\@tempboxa\hbox{\normalfont\footnotesize {#1.}\nobreakspace\nobreakspace #2}%
\ifdim \wd\@tempboxa >\hsize%
\setbox\@tempboxa\hbox{\normalfont\footnotesize {#1.}\nobreakspace\nobreakspace}%
\parbox[t]{\hsize}{\normalfont\footnotesize\noindent\unhbox\@tempboxa#2}%
\else%
\ifCLASSOPTIONconference \hbox to\hsize{\normalfont\footnotesize\hfil\box\@tempboxa\hfil}%
\else \hbox to\hsize{\normalfont\footnotesize\box\@tempboxa\hfil}%
\fi\fi\fi}
\makeatother

\newcommand{\stepnum}[1]{%
\tikz[baseline=(char.base)]{
  \node[shape=circle, fill=black, text=white, inner sep=1pt] (char) {\small #1};
}}

\setlist{nosep,leftmargin=*}

\lstdefinelanguage{Verilog}{
  morekeywords={module,endmodule,input,output,reg,wire,always,posedge,negedge,
    if,else,begin,end,assign,parameter,localparam,integer,genvar,generate,
    endgenerate,case,endcase,default,initial,forever},
  sensitive=true,
  morecomment=[l]{//},
  morecomment=[s]{/*}{*/},
  morestring=[b]",
}
\definecolor{fgcggoodbg}{HTML}{F1FAF4}
\definecolor{fgcggoodrule}{HTML}{2E7D32}
\definecolor{fgcgbadbg}{HTML}{FFF1F0}
\definecolor{fgcgbadrule}{HTML}{B3261E}
\DeclareRobustCommand{\figstep}[1]{\raisebox{.15ex}{\textcircled{\scriptsize #1}}}
\title{\autogate: Automated Clock Gating via Toggling-Aware LLM-based RTL Rewriting}

\author{
\IEEEauthorblockN{
Yiting Wang\IEEEauthorrefmark{1}\IEEEauthorrefmark{2},
Chenhui Deng\IEEEauthorrefmark{3},
Chia-Tung Ho\IEEEauthorrefmark{3},
Yanqing Zhang\IEEEauthorrefmark{3},
Zhuo Feng\IEEEauthorrefmark{3},\\
Cunxi Yu\IEEEauthorrefmark{2}\IEEEauthorrefmark{3},
Ang Li\IEEEauthorrefmark{2},
Gang Qu\IEEEauthorrefmark{2},
Brucek Khailany\IEEEauthorrefmark{3}
}
\IEEEauthorblockA{\IEEEauthorrefmark{2}University of Maryland, College Park, \IEEEauthorrefmark{3}NVIDIA\\
}
\IEEEauthorblockA{\{ywang144, angliece, gangqu\}@umd.edu, 
 \{cdeng, chiatungh, yanqingz, zhuof, cunxiy, bkhailany\}@nvidia.com}
\thanks{\IEEEauthorrefmark{1}This work was conducted during an internship at NVIDIA.}
}

\begin{document}
\maketitle

\input{sections/abstract}
\input{sections/introduction}
\input{sections/background}

\input{sections/methodology}
\input{sections/results}

\input{sections/conclusion}

\balance
\bibliographystyle{IEEEtran}
\bibliography{paper}

\end{document}

%% file: sections/abstract.tex
\begin{abstract}

Fine-grain clock gating (FGCG) is among the most effective techniques for reducing dynamic power, yet current FGCG optimization flows remain largely manual. Recent LLM-based RTL optimization approaches remain limited by two key drawbacks: (1) the inability to process long waveform traces spanning millions of cycles, and (2) the difficulty of scaling optimization to large hierarchical codebases while preserving correctness.
In this work, we present \textbf{\autogate}, the first agentic framework for industry-grade RTL power optimization, enabling workload-aware clock-gating optimization across large hierarchical codebases. \autogate introduces a Machine Learning (ML)–LLM co-design that bridges waveform-level analysis and RTL rewriting. Specifically, we design an ML-based clustering algorithm that distills raw toggling traces into compact, structured representations that guide LLM-based RTL rewriting. This enables accurate identification and application of clock-gating opportunities without requiring LLMs to directly process raw waveform data. To enhance scalability, \autogate employs a hierarchical multi-agent architecture that decomposes large designs into independently optimizable modules, enabling coordinated optimization across deep design hierarchies.
We evaluate \autogate on a diverse set of designs ranging from small RTL designs to large industrial-grade codebases. 
Experimental results show that \autogate consistently reduces dynamic power relative to baselines.
Across the small-design suite, \autogate reduces dynamic power by 49.31\% on average. On industry-scale designs, it achieves 19.34\% and 7.96\% dynamic power reductions on NVDLA~\cite{nvdla_hw} and BlackParrot~\cite{petrisko2020blackparrot}, respectively, and up to 6.86\% on highly optimized proprietary production designs.

\end{abstract}


%% file: sections/introduction.tex
\section{Introduction}
\label{sec:intro}

Power consumption is a first-order design constraint in modern very-large-scale integration (VLSI) systems on a chip. 
Dynamic power is driven by switching on clock networks, data paths, and interconnects.
Fine-grain clock gating (FGCG) gates small groups of flip-flops to reduce dynamic power consumption~\cite{benini1999survey}.
Today's commercial synthesis tools support automatic clock-gating inference at the register transfer level (RTL) by exploiting clock-gating opportunities that are explicitly exposed by the RTL coding style. Consequently, many workload-dependent FGCG opportunities cannot be discovered by these synthesis tools before RTL restructuring. Such RTL rewriting is overwhelmingly
manual, driven by rigid heuristics and the design team's experience. Workload-aware automated FGCG optimization remains largely unexplored.

Recent large language models (LLMs) have demonstrated strong capabilities in RTL code understanding, generation, and optimization~\cite{deng2025scalertl, deng2026ace, blocklove2023chipchat, wang2025symrtlo,yao2024rtlrewriter, ping2026poet}, making LLMs promising candidates for automated RTL rewriting to improve FGCG. However, their practical adoption is restricted by the following two limitations.


\textbf{First, FGCG requires the analysis of long \textit{waveform} that exceeds the capabilities of current LLMs}. Effective FGCG relies on cycle-accurate toggling behavior that is tightly coupled with input stimuli and workload characteristics, requiring analysis of waveform traces spanning millions of clock cycles. Such traces translate into extremely long contexts, far beyond what LLMs can process accurately and reliably, making it difficult to reason and discover the patterns that might lead to dynamic power bottlenecks. Consequently, raw waveform data must first be analyzed and summarized into compact representations before it can be processed in any LLM-based flow. Meanwhile, FGCG clustering algorithms, from coarse toggle-rate-based grouping to sequence-level clustering methods~\cite{gluzer2016probability, park2023ml, le2015stability}, can identify candidate flip-flop groups, but rewriting/translating these into correct and efficient RTL is challenging because of the need for deep RTL understanding and careful handling of control and data dependencies. This raises an open question: how can LLM-based RTL optimization be effectively integrated with specialized Machine-Learning (ML)-driven waveform analysis to enable accurate clock-gating decisions without requiring LLMs to directly process raw switching traces?





\textbf{Second, existing LLM-based RTL optimization approaches do not scale to large and hierarchical industrial-grade designs.} Prior LLM-based RTL optimization studies \cite{yao2024rtlrewriter, wang2025symrtlo, ping2026poet} have primarily focused on small, single-module designs with only a few hundred lines of code, whereas industrial designs consist of deeply hierarchical codebases with many interacting modules and complex cross-module dependencies, as demonstrated in Table \ref{tab:motivation_small_large}. In such settings, a single-agent, single-context approach does not scale, as effective optimization requires coordinated reasoning across the entire design hierarchy. 
The challenge is further compounded by large SoC designs typically providing only top-level testbenches. Because FGCG relies on module-level switching activity, practical optimization must also obtain module-level behavior from full-design simulations.

\vspace{-5pt}
\begin{table}[h]
\centering
\caption{Motivational comparison of small and large RTL designs. Values show dynamic-power change relative to baseline.}
\vspace{-5pt}
\label{tab:motivation_small_large}
\small
\renewcommand{\arraystretch}{0.6}
\scriptsize
\begin{tabular*}{\columnwidth}{@{\extracolsep{\fill}}lrrrr@{}}
\toprule
\textbf{Design} & \textbf{\# Files} & \textbf{POET~\cite{ping2026poet}} & \textbf{ROVER~\cite{coward2024combining}} & \textbf{\autogate} \\
\midrule
booth\_mult & 1   & \textcolor{green!60!black}{\(\downarrow\)~$-$85.52\%} & \textcolor{green!60!black}{\(\downarrow\)~$-$79.72\%} & \textcolor{green!60!black}{\(\downarrow\)~$-$86.21\%} \\
NVDLA total & 135 & +0.09\% & $-$0.01\% & \textcolor{green!60!black}{\(\downarrow\)~$-$19.34\%} \\
\bottomrule
\end{tabular*}
\vspace{-10pt}
\end{table}




\smallskip\noindent\textbf{Our Approach.} We present \textbf{\autogate}, a workload-aware power optimization framework for industrial-scale RTL codebases through hierarchical decomposition, activity-guided optimization, formal verification, and automated FGCG rewriting. \autogate rewrites RTL to expose additional gating opportunities, enabling synthesis to infer more effective clock gating.


\autogate addresses the \textbf{limitations of LLMs in processing long waveform traces} through an adaptive switching-activity analysis. Specifically, we transform raw simulation traces into a hierarchy of clock-gating candidates using toggle-aware pre-filtering, automatic threshold discovery, and multi-threshold stability clustering. The resulting gating candidates are then distilled into a compact, structured representation that guides LLM-based rewriting, enabling the LLM to expose identified clock-gating opportunities while preserving control and data dependencies, thereby bridging waveform-level activity analysis and RTL optimization.
Notably, when improving gating efficiency, the LLM may merge existing clock gates in the original design. As a result, area reduction may occur as a byproduct of clock-gate merging.



To scale up to \textbf{industrial-grade RTL codebases}, \autogate uses a \textit{divide-and-conquer} strategy that decomposes complex hierarchical designs into independently optimizable modules, enabling scalable and coordinated optimization across the entire design. We use a multi-agent architecture in which an orchestrator LLM automatically identifies the design hierarchy, partitions the codebase into manageable units, and leverages per-module switching activity analysis to guide optimization. This approach enables \autogate to scale effectively to production-level designs while preserving cross-module correctness and optimization consistency.

 We conduct extensive experiments on 10 designs, spanning small single modules, industrial-grade designs, and highly optimized proprietary production designs. Our results demonstrate that AutoGate consistently outperforms competitive baselines in terms of dynamic power reduction.

\smallskip\noindent\textbf{Our contributions are as follows}:
\begin{enumerate}
\item We propose \autogate,  the first agentic framework for industrial-grade RTL power optimization, enabling workload-aware FGCG across large hierarchical codebases. 

\item We address the limitations of LLMs in precisely understanding long-context waveforms by co-designing an ML-based toggling analysis method to guide LLM decisions. This combines the strengths of ML-based long sequence processing and LLM-based RTL reasoning to improve FGCG effectiveness.

\item We demonstrate that \autogate exposes clock-gating opportunities beyond those identified by a commercial synthesis flow with aggressive automatic clock-gating inference enabled. Across small benchmarks, \autogate achieves an average dynamic-power reduction of 49.31\% with only 0.15\% average area overhead.



\item We overcome the scalability limitations of LLM-based RTL optimization through divide-and-conquer optimization of large hierarchical codebases with complex cross-module dependencies. On large industrial-grade designs, \autogate achieves average dynamic-power reductions of 19.34\% on NVDLA with a 3.46\% average area reduction across partitions, 7.96\% on BlackParrot with 0.04\% area overhead, and up to 6.86\% on highly optimized proprietary production designs with at most 1\% area overhead.


\end{enumerate}

%% file: sections/background.tex
\section{Background}
\label{sec:background}

\subsection{Dynamic Power and Clock Gating}
Dynamic power is a dominant component of power consumption in modern digital designs. It is commonly modeled as
\begin{equation*}
    P_{\mathrm{dyn}} = \alpha \cdot C_L \cdot V_{DD}^{2} \cdot f_{\mathrm{clk}},
\end{equation*}
where $\alpha$ is the switching activity factor, $C_L$ is the load
capacitance, $V_{DD}$ is the supply voltage, and $f_{\mathrm{clk}}$ is
the clock frequency. At the RTL stage, the supply voltage, technology
capacitance, and target clock frequency are usually fixed by design
constraints. Therefore, reducing unnecessary switching activity is one
of the most effective methods for dynamic power optimization.

Fine-grain clock gating (FGCG) reduces unnecessary switching by inserting
integrated clock-gating (ICG) cells to small groups of flip-flops. When
the registers in a group do not need to update, the ICG suppresses the
clock, reducing switching in both the clock path and the flip-flops.
However, FGCG is not free: each additional ICG introduces area, control
logic, and clock-tree overhead. Effective FGCG therefore requires
selecting register groups whose idle cycles align well enough to justify
the added gating logic.

\subsection{RTL Coding Style of inserting FGCG}
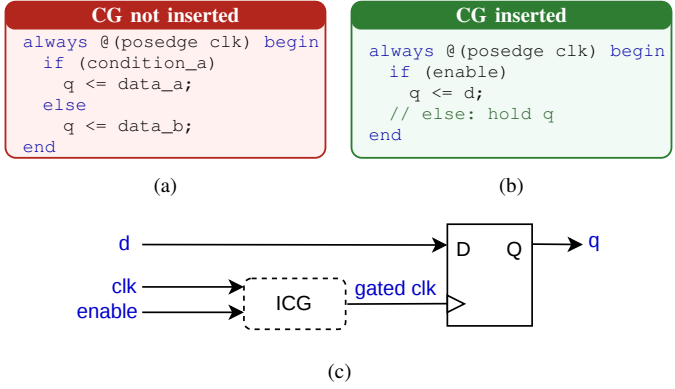
\begin{figure}[h]
\centering
\resizebox{\columnwidth}{!}{%
\begin{tikzpicture}[
  panel/.style={rounded corners=4pt, line width=0.7pt, minimum width=4.3cm, minimum height=2.1cm, inner sep=0pt},
  title/.style={rounded corners=3.5pt, minimum width=4.3cm, minimum height=0.38cm, inner sep=0pt, font=\bfseries\footnotesize, text=white, align=center},
  code/.style={anchor=north west, align=left, font=\ttfamily\scriptsize, text=black},
  subcap/.style={anchor=north, font=\footnotesize}
]
\node[panel, draw=fgcgbadrule, fill=fgcgbadbg, anchor=north west] (bad) at (0,0) {};
\node[title, fill=fgcgbadrule, anchor=north] at (bad.north) {CG not inserted};
\node[code] at ([xshift=0.12cm,yshift=-0.34cm]bad.north west) {
\textcolor{blue!65!black}{always} @(posedge clk) \textcolor{blue!65!black}{begin}\\
\hspace*{1.1em}\textcolor{blue!65!black}{if} (condition\_a)\\
\hspace*{2.2em}q \textless= data\_a;\\
\hspace*{1.1em}\textcolor{blue!65!black}{else}\\
\hspace*{2.2em}q \textless= data\_b;\\
\textcolor{blue!65!black}{end}
};
\node[subcap] at ([yshift=-0.13cm]bad.south) {(a)};
\node[panel, draw=fgcggoodrule, fill=fgcggoodbg, anchor=north west] (good) at (4.65,0) {};
\node[title, fill=fgcggoodrule, anchor=north] at (good.north) {CG inserted};
\node[code] at ([xshift=0.12cm,yshift=-0.46cm]good.north west) {
\textcolor{blue!65!black}{always} @(posedge clk) \textcolor{blue!65!black}{begin}\\
\hspace*{1.1em}\textcolor{blue!65!black}{if} (enable)\\
\hspace*{2.2em}q \textless= d;\\
\hspace*{1.1em}\textcolor{fgcggoodrule!75!black}{// else: hold q}\\
\textcolor{blue!65!black}{end}
};
\node[subcap] at ([yshift=-0.13cm]good.south) {(b)};
\end{tikzpicture}%
}

\vspace{0.4em}

\includegraphics[width=0.85\columnwidth]{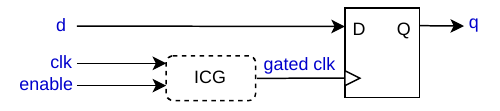}\\[0.2em]
{\footnotesize (c)}

\vspace{-0.3em}
\caption{RTL coding styles for synthesis-inferred FGCG and the corresponding gated-register schematic.}
\label{fig:fgcg_rtl_style}
\end{figure}

In standard digital design flows, logic synthesis tools infer FGCG from RTL coding patterns rather than from manually instantiated ICG cells. Fig.~\ref{fig:fgcg_rtl_style}
compares two common coding styles: the right style exposes an explicit
hold condition, while the left style assigns the register on every cycle.

In Fig.~\ref{fig:fgcg_rtl_style}(a), \texttt{q} is assigned every cycle and therefore does not expose a clock-gating opportunity. In contrast, the explicit hold condition in Fig.~\ref{fig:fgcg_rtl_style}(b) enables synthesis to infer a clock gate, whose hardware implementation is shown in Fig.~\ref{fig:fgcg_rtl_style}(c). FGCG-oriented RTL rewriting therefore transforms functionally equivalent code into forms that expose register hold behavior and enable clock-gating inference.




\vspace{-5pt}
\subsection{Related Work}

Recent LLM-based hardware design methods have demonstrated promising capabilities in Verilog generation~\cite{deng2025scalertl, deng2026ace}, chip design assistance~\cite{blocklove2023chipchat}, and RTL optimization through LLM-guided rewriting~\cite{yao2024rtlrewriter}, e-graph rewriting~\cite{zhang2025aspen}, symbolic reasoning~\cite{wang2025symrtlo}, and evolutionary search~\cite{ping2026poet}. However, existing methods neither address the workload-dependent nature of automated FGCG nor provide a scalable solution for industrial-scale RTL repositories.
First, FGCG is workload-dependent and requires processing cycle-level switching activity, while realistic traces can span millions of cycles and exceed practical LLM context limits. Second, most RTL optimization methods target small or single-module designs and do not scale to hierarchical industrial-grade RTL codebases with cross-module dependencies. Non-LLM approaches face similar limitations: rule-based methods~\cite{coward2024combining} rely on predefined rewrite patterns with limited generalization capability, while clustering-based flows~\cite{gluzer2016probability, park2023ml, le2015stability} require manual tuning and expert interpretation to translate clustering results into effective RTL rewrites. These limitations motivate a framework that combines workload-aware FGCG analysis with hierarchy-aware LLM-based RTL rewriting.
Table~\ref{tab:related_comparison} compares the capabilities of existing RTL optimization frameworks with \autogate.


\begin{table}[t]
\centering
\caption{Comparison of LLM-based RTL optimization methods. Power Aware indicates explicit power optimization, Clock Gating indicates support for clock-gating transformations, Agent Based indicates an agentic workflow, Repo Level indicates support for large, multi-file RTL codebases, and Custom Algo. indicates custom algorithm-guided rewriting. }

\label{tab:related_comparison}
\small
\renewcommand{\arraystretch}{0.6}
\scriptsize
\setlength{\tabcolsep}{7pt}
\begin{tabular}{lccccc}
\toprule
\textbf{Method} & \textbf{\makecell{Power\\Aware}} & \textbf{\makecell{Clock\\Gating}} & \textbf{\makecell{Agent\\Based}} & \textbf{\makecell{Repo\\Level}} & \textbf{\makecell{Custom\\Algo.}} \\
\midrule
RTLRewriter~\cite{yao2024rtlrewriter}  & \cmark & \xmark & \cmark & \xmark & \xmark \\
ROVER~\cite{coward2024combining}       & \cmark & \cmark & \xmark & \xmark & \cmark \\
SymRTLO~\cite{wang2025symrtlo}         & \cmark & \xmark & \cmark & \xmark & \cmark \\
ASPEN~\cite{zhang2025aspen}            & \xmark & \xmark & \cmark & \xmark & \cmark \\
POET~\cite{ping2026poet}               & \cmark & \xmark & \cmark & \xmark & \xmark \\
\textbf{\autogate (Ours)}               & \cmark & \cmark & \cmark & \cmark & \cmark \\
\bottomrule
\end{tabular}
\vspace{-15pt}
\end{table}

\vspace{-5pt}
\subsection{Motivational Experiment}
\label{subsec:motivation_exp}

To demonstrate the limitations in existing RTL optimization methods in hierarchical RTL codebases, we have performed a motivational experiment.
 Table~\ref{tab:motivation_small_large}
shows this gap using one single-module benchmark and one large RTL design.
On the single module booth multiplier design, both LLM-based method (\textit{POET}~\cite{ping2026poet}) and non LLM-based method (\textit{ROVER}~\cite{coward2024combining}) significantly reduce dynamic
power. However, when the dynamic power of the reported NVDLA partitions is
aggregated under the same workload, both methods provide nearly no
improvement. This motivates a scalable, hierarchy-aware optimization flow
for large hierarchical RTL codebases.

%% file: sections/methodology.tex

\begin{figure*}[!h]
\centering
\includegraphics[width=0.95\textwidth]{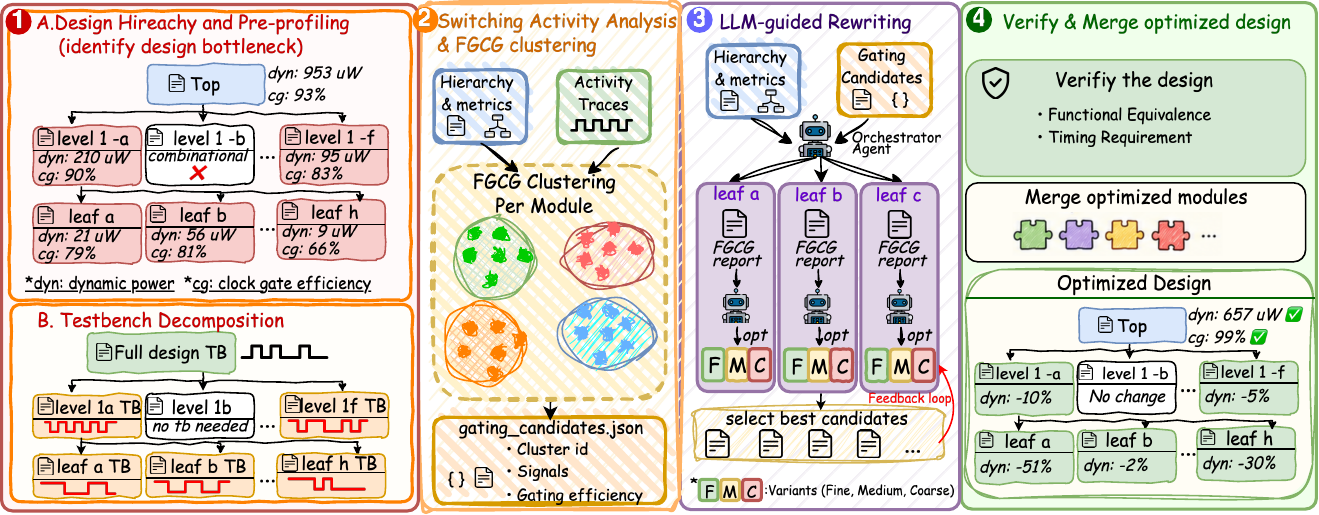}
\caption{\autogate framework overview. \figstep{1} identifies optimization targets and extracts module-level workloads; \figstep{2} generates FGCG candidates from switching activity; \figstep{3} performs parallel LLM-guided RTL rewriting; and \figstep{4} verifies and selects QoR-improving candidates.}

\label{fig:framework}
\end{figure*}

\vspace{-5pt}
\section{Methodology}
\label{sec:methodology}

\autogate framework consists of four stages as shown in Fig.~\ref{fig:framework}, and we will elaborate each of them next.

\subsection{Stage 1: Hierarchy Detection and Pre-Profiling}
\label{sec:stage1}

Large RTL codebases typically exhibit deep and complex design hierarchies with massive code volume, often exceeding the context window of LLMs and making holistic understanding impractical. Furthermore, synthesizing and performing power analysis on the entire codebase is computationally expensive, often requiring hours to days, which makes iterative optimization with power feedback very challenging at the full-design level. To address these challenges, \autogate adopts a divide-and-conquer strategy that decomposes the design into modules, identifies power-critical components, and applies targeted optimization to each module independently, as shown in Fig.~\ref{fig:framework} Stage \figstep{1}.

\autogate begins by constructing an explicit design hierarchy from the RTL source tree, where the orchestrator parses module definitions and instantiations to resolve parent–child relationships and identify independently optimizable modules. It then traverses this hierarchy using depth-first search to perform per-module pre-profiling, recording structural features such as module paths, sequential logic regions, and register counts.

To incorporate power information, \autogate maps synthesized gate-level instances back to their originating RTL files and source lines, and aggregates gate-level power estimates by RTL region and logic category. This is combined with synthesis and power metrics, including total and dynamic power, clock-gating coverage and efficiency, flop count, and ICG count. 

The resulting structured \emph{power bottleneck report} ranks modules by optimization value and captures source-level power attribution, clock-gating headroom, and downstream impact, allowing \autogate to focus only on promising candidates and avoid unnecessary rewriting. By following this divide-and-conquer strategy, \autogate requires full-design synthesis and power analysis only once.

FGCG analysis is a key component of clock-gating candidate selection and requires per-module switching traces. However, large RTL projects typically provide only top-level or subsystem-level testbenches, and manually writing standalone testbenches for every internal module would undermine automation. \autogate addresses this by extracting module-level testbenches directly from full-design simulation traces for each bottleneck module. It parses module interfaces, extracts scoped waveform data, and generates standalone testbenches that replay real workload inputs cycle by cycle. This enables accurate per-module switching analysis while preserving realistic system behavior, avoiding reliance on synthetic stimuli.

\subsection{Stage 2: Adaptive Stability-Based FGCG Exploration}
\label{sec:clustering}

\begin{figure*}[!t]
\centering
\vspace{-10pt}
\includegraphics[width=0.95\textwidth]{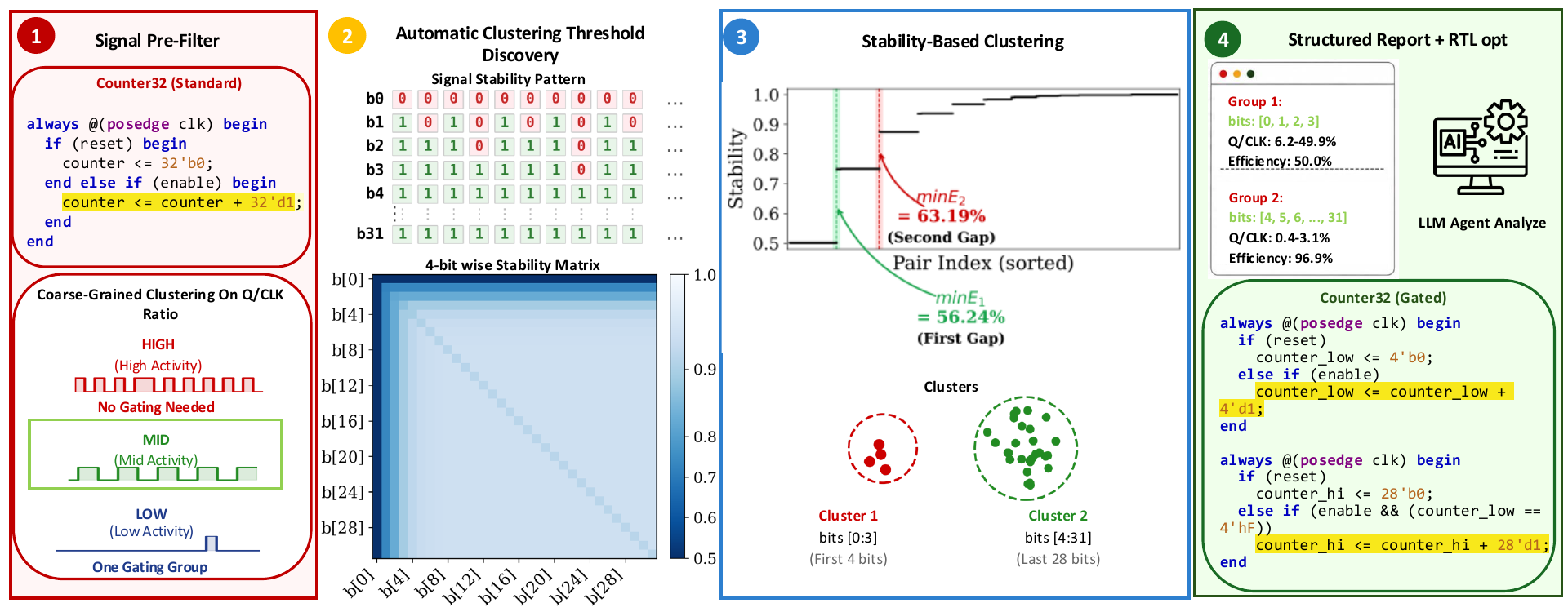}

\caption{Adaptive Stability FGCG clustering flow.}
\label{fig:fgcg_flow}
\vspace{-15pt}
\end{figure*}

Unlike prior FGCG methods that produce a single clustering result, \autogate generates a hierarchy of candidate clusterings spanning different power-area tradeoffs. As shown in Fig.~\ref{fig:framework} Stage \figstep{2} and detailed in Fig.~\ref{fig:fgcg_flow}, RTL switching traces from the extracted testbench are processed through four steps: \stepnum{1} toggle-aware pre-filtering partitions registers into activity bands; \stepnum{2} switching traces are converted into binary stability patterns and are then assembled into the stability matrix for automatic threshold discovery; \stepnum{3} stability-based clustering forms candidate FGCG groups; and \stepnum{4} structured optimization reports are generated for the LLM agents.


\smallskip\noindent\textbf{Stability Patterns and Group Efficiency.\cite{le2015stability}}

Let $\mathcal{R}=\{r_1,\ldots,r_N\}$ be the register set under analysis, and $T$ be the number of observed clock edges.
A candidate gating group $G\subseteq\mathcal{R}$ is the set of registers proposed to share one inferred ICG. 
The \textit{stability pattern} of register $a$ is a binary sequence
$\mathrm{SP}_a \in \{0,1\}^{T}$, where $\mathrm{SP}_a[k] = 1$ indicates
that register $a$ remains unchanged at clock edge $k$, and
$\mathrm{SP}_a[k] = 0$ otherwise.
Since that ICG can be disabled only when all registers in $G$ are stable, the group stability is the element-wise logical AND $\text{SP}_G[k]=\bigwedge_{a\in G}\text{SP}_a[k]$.
The group \emph{gating efficiency} $E(G)=\frac{1}{T}\sum_{k=1}^{T}\text{SP}_G[k]$ is therefore the fraction of cycles in which the whole group can safely gate its clock.

Clustering starts from singleton groups and repeatedly merges the closest pair of current clusters $C_i,C_j\subseteq\mathcal{R}$ under $d(C_i,C_j)=\|\text{SP}_{C_i}-\text{SP}_{C_j}\|_2$, accepting a merged cluster $M=C_i\cup C_j$ only when its ANDed stability vector satisfies $E(M)\geq\texttt{minE}$.


\vspace{-5pt}
\smallskip\noindent\textbf{\stepnum{1} Toggling-aware Pre-Filter.}

To avoid applying unnecessary clock gating to flops with little optimization benefit, \autogate first filters registers based on their toggle rates. Registers with toggle rates close to 0 or 1 are unlikely to benefit from gating and are therefore excluded. Specifically, before stability clustering, \autogate partitions registers in each module into low, middle, and high toggle-rate bands using their measured Q/CLK ratio under the target workload, with thresholds of 3\% and 95\% for the low and high bands, respectively. Stability clustering is then applied only to the middle band to reduce computational cost and focus on the most promising candidates. If the middle band is empty, a fallback mode is triggered that includes all registers. These threshold values are fixed across all experiments; systematic tuning of these parameters is out of the scope of this work and is left for future work.
  
\smallskip\noindent\textbf{\stepnum{2} Automatic Threshold Selection.}
The stability-based FGCG clustering algorithm \cite{le2015stability} depends on a manually chosen minimum group gating-efficiency threshold $\texttt{minE}$. This threshold is fragile: too low merges unrelated registers into one gate, while too high produces single-register clusters that waste area. \autogate removes this manual tuning by computing a fixed-size \emph{group-wise efficiency spectrum}.


For a candidate group size $b$, we set $b$ to the minimum clock-gating bit-width. Unless otherwise specified, $b=4$, matching the default setting of the commercial synthesis tool. The efficiency spectrum evaluates every $b$-register candidate group:
$\mathcal{E}_b=\{E(H)\mid H\subseteq\mathcal{R},\,|H|=b\}$, where
$E(H)=\frac{1}{T}\sum_{k=1}^{T}\min_{a\in H}\mathrm{SP}_a[k]$.
 $\mathcal{R}$ denotes the register set under analysis. The algorithm sorts the $\binom{N}{b}$ group efficiencies, identifies the largest consecutive gaps, and uses the corresponding gap midpoints as candidate $\texttt{minE}$ values. For large register sets, we sample candidate groups to identify the threshold gaps and avoid combinatorial explosion.

\smallskip\noindent\textbf{\stepnum{3} Multi-Threshold Clustering.}
To explore the power-area trade-offs across different clustering granularity and identify the most balanced rewrite options,
\autogate runs the clustering algorithm at each detected group gating-efficiency threshold to generate a hierarchy of candidate FGCG assignments. The finest level creates more ICGs and maximizes gating opportunity; coarser levels reduce ICG count and area overhead. An adaptive sweep probes around each threshold with scale-proportional offsets to provide additional actionable multi-register groups, giving the LLM agents a richer set of rewrite options.

\smallskip\noindent\textbf{\stepnum{4} Gating-Candidate Report.} For every high-impact module, the report lists inefficient ICGs, the registers responsible for wasted toggles, recommended FGCG cluster assignments, and RTL regions where rewrites should be applied. The report is structured to be concise for LLM consumption: it names the relevant signals, ranks them by power impact, and explains the gating pattern that the rewrite should expose.

\vspace{-5pt}
\subsection{Stage 3: LLM-Guided Parallel Rewriting}
\label{sec:agentic}

\vspace{-5pt}
To enable scalable RTL optimization, \autogate adopts an agentic framework that decomposes rewriting into coordinated module-level transformations. Using the hierarchy and clustering reports from Stages~1 and 2, an orchestrator assigns modules to sub-agents and schedules optimization bottom-up through the design hierarchy. Leaf modules are optimized first, parent modules only after child modules optimized, and independent modules are optimized in parallel.

Each module is paired with a dedicated rewrite agent that operates only on the target RTL, guided by its power-bottleneck report, gating-candidate clusters under different group gating-efficiency thresholds, and a library of clock-gating rewrite templates. These templates capture common RTL transformation patterns that expose implicit idle behavior and enable synthesis-inferred clock gating. Representative patterns include: (1) \emph{enable/hold extraction}, which converts unconditional assignments into conditional updates to reveal register hold conditions (e.g., rewriting \texttt{q <= d;} into \texttt{if (en) q <= d;}); (2) \emph{value-change-based gating}, which guards updates based on data changes (e.g., \texttt{if (q != d) q <= d;}); (3) \emph{control-signal factoring}, which lifts shared enable conditions across multiple registers to form a common gating signal; and (4) \emph{register decomposition}, which splits wide registers into sub-registers with hierarchical enables to increase gating opportunities. Combined with clustering guidance, these templates enable functionally equivalent RTL rewrites that are more amenable to FGCG insertion.


For each module, the agent generates fine-, medium-, and coarse-grain variants. Fine-grain variants maximize gating opportunities, medium-grain variants merge nearby clusters to reduce ICG count, and coarse-grain variants retain only the highest-efficiency groups to balance power and area. Only the target module for each rewrite agent is synthesized and evaluated for power, avoiding full-design recompilation.

A reflection agent evaluates generated variants, identifies the causes of power improvements or regressions, and distills them into concise lessons stored in shared global and module-specific playbooks. Guided by both power feedback and accumulated playbook knowledge, rewrite agents iteratively refine their transformations to improve optimization quality.

\vspace{-5pt}
\subsection{Stage 4: Formal Verification and QoR-Aware Merge}
\label{sec:qor}

\autogate validates each candidate through syntax checking, simulation, commercial formal equivalence checking, and timing validation. Only candidates that preserve functionality and timing closure proceed to QoR evaluation.
Each surviving candidate is synthesized and profiled for dynamic power, area, clock-gating metrics, and timing, yielding a QoR profile containing $\Delta P_{\text{dyn}}$, $\Delta A$, clock-gating coverage and efficiency, ICG count, WNS, and TNS. The final candidate is selected to maximize dynamic-power reduction under predefined area and timing constraints; otherwise, the original module is retained.
Accepted patches are merged hierarchically and revalidated at the parent level. 

%% file: sections/results.tex

\section{Experimental Results}
\label{sec:results}


\begin{table}[t]
\centering
\caption{Benchmark design summary. Lines of code, cell counts, and token counts are reported for the original unoptimized RTL design.}
\vspace{-5pt}
\label{tab:benchmarks}
\renewcommand{\arraystretch}{0.8}
\scriptsize
\setlength{\tabcolsep}{5pt}
\begin{tabularx}{0.95\columnwidth}{@{}Xrrrr@{}}
\toprule
\textbf{Design} & \textbf{Files} & \textbf{Lines} & \textbf{Cells} & \textbf{Tokens} \\
\specialrule{0.14em}{0pt}{0pt}
\multicolumn{5}{@{}l}{\textbf{\textit{Small designs}}} \\
\midrule
counter32            & 1  & 17    & 101     & 85   \\
booth\_mult          & 1  & 81    & 806     & 792  \\
divide               & 1  & 121   & 757     & 842  \\
aes\_cipher          & 1  & 932   & 103,582 & 11K  \\
deframer             & 20 & 243   & 148     & 2.4K \\
mrisc                & 7  & 1,516 & 2,039   & 15K  \\
\specialrule{0.14em}{0pt}{0pt}
\multicolumn{5}{@{}l}{\textbf{\textit{Large designs}}} \\
\midrule
NVDLA total          & 135 & 103,079   & 1,757,639 & 1.0M \\
\hspace{0.5em}\textcolor{brown}{\textbf{$\vert$----}}\;Partition\_C & 43  & 54,499  & 1,654,571 & 545K \\
\hspace{0.5em}\textcolor{brown}{\textbf{$\vert$----}}\;Partition\_M & 12  & 5,581   & 14,975    & 56K  \\
\hspace{0.5em}\textcolor{brown}{\textbf{$\vert$----}}\;Partition\_A & 16  & 8,140   & 14,915    & 81K  \\
\hspace{0.5em}\textcolor{brown}{\textbf{$\vert$----}}\;Partition\_P & 64  & 34,859  & 73,178    & 349K \\
\midrule
BlackParrot         & 77  & 1,071,596 & 1,140,814 & 30.6M \\
\bottomrule
\end{tabularx}
\vspace{-15pt}
\end{table}

\begin{table*}[!h]
\centering
\caption{\autogate results on small benchmark designs. ``Base'' denotes the original design; ``POET'' denotes the POET~\cite{ping2026poet} result; ``ROVER'' denotes the ROVER~\cite{coward2024combining} result; ``\autogate'' denotes the optimized design. CG~Cov = clock-gating coverage; CG~Eff = clock-gating efficiency.}
\vspace{-5pt}
\label{tab:cvdp}
\renewcommand{\arraystretch}{0.6}
\scriptsize

\setlength{\tabcolsep}{6pt}
\setlength{\aboverulesep}{0pt}
\setlength{\belowrulesep}{0pt}
\footnotesize
\begin{tabular}{l r r r r r r r c c c}

\toprule
\textbf{Design} & \textbf{\makecell{Total Pwr\\(uW)}} & \textbf{\makecell{Dyn Pwr\\(uW)}} & \textbf{Dyn $\Delta$} & \textbf{\makecell{Area\\(um\textsuperscript{2})}} & \textbf{Area $\Delta$} & \textbf{CG Cov} & \textbf{CG Eff} & \textbf{ICGs} & \textbf{\makecell{WNS\\(ns)}} & \textbf{\makecell{Timing\\Pass}} \\
\midrule
\rowcolor{basegray} counter32\_base     & 55.76  & 53.23  & ---     & 7.63   & ---      & 100\% & 22.90\% & 1 & 0.45 & \cmark \\
counter32\_POET     & 52.37  & 49.02  & $-$7.91\%   & 7.63   & 0.00\%    & 100\% & 30.60\% & 1 & 0.44 & \cmark \\
counter32\_ROVER    & 56.78  & 54.22  & +1.86\%    & 8.17   & +7.08\%  & 100\% & 24.50\% & 1 & 0.27 & \cmark \\
\rowcolor{autogategreen} counter32\_\autogate & 26.69  & \textbf{23.78}  & \textbf{$-$55.33\%}  & 7.75   & +1.57\%   & 100\% & 82.80\% & 2 & 0.00 & \cmark \\
\midrule
\rowcolor{basegray} booth\_mult\_base   & 708.90 & 683.69 & ---     & 95.50  & ---      & 9.30\%   & 8.30\%  & 1 & 0.00 & \cmark \\
booth\_mult\_POET   & 119.20 & 99.00  & $-$85.52\%  & 62.33  & $-$34.73\% & 99.40\%  & 86.00\% & 6 & 0.00 & \cmark \\
booth\_mult\_ROVER  & 169.20 & 138.66 & $-$79.72\% & 95.99  & +0.51\%  & 98.60\%  & 88.00\% & 5 & 0.00 & \cmark \\
\rowcolor{autogategreen} booth\_mult\_\autogate & 114.40 & \textbf{94.30}  & \textbf{$-$86.21\%} & 62.11  & $-$34.96\% & 97.00\%  & 87.00\% & 5 & 0.00 & \cmark \\
\midrule
\rowcolor{basegray} divide\_base        & 672.80 & 645.20 & ---     & 83.93  & ---      & 100\% & 15.40\% & 1 & 0.00 & \cmark \\
divide\_POET        & 671.60 & 644.40 & $-$0.12\%   & 83.13  & $-$0.95\% & 100\% & 15.40\% & 1 & 0.00 & \cmark \\
divide\_ROVER       & 699.00 & 667.64 & +3.48\%    & 100.61 & +19.87\% & 100\% & 15.40\% & 1 & 0.00 & \cmark \\
\rowcolor{autogategreen} divide\_\autogate    & 302.60 & \textbf{265.40} & \textbf{$-$58.87\%}  & 103.63 & +23.47\%  & 96.40\%  & 81.50\% & 31 & 0.00 & \cmark \\
\midrule
\rowcolor{basegray} aes\_base           & 12383.00 & 8019.00 & ---        & 12463.42 & ---      & 99.97\% & 98.31\% & 361 & 0.00 & \cmark \\
aes\_POET           & 10791.00 & 6653.00 & $-$17.03\%  & 12467.26 & +0.03\%  & 99.96\% & 98.30\% & 360 & 0.00 & \cmark \\
aes\_ROVER          & 12418.00 & 8040.00 & +0.26\%     & 12462.21 & $-$0.01\% & 99.96\% & 98.30\% & 360 & 0.00 & \cmark \\
\rowcolor{autogategreen} aes\_\autogate      & 10144.00 & \textbf{6056.00} & \textbf{$-$24.48\%} & 12515.56 & +0.42\%  & 99.95\% & 98.40\% & 289 & 0.00 & \cmark \\
\midrule
\rowcolor{basegray} deframer\_base      & 99.17  & 93.85  & ---     & 13.57  & ---      & 100\% & 49.40\% & 6 & 0.00 & \cmark \\
deframer\_POET      & 88.03  & 82.81  & $-$11.76\% & 13.83  & +1.92\%  & 100\% & 57.00\% & 8 & 0.00 & \cmark \\
deframer\_ROVER     & 97.18  & 92.05  & $-$1.92\%  & 13.76  & +1.40\%  & 100\% & 49.40\% & 6 & 0.00 & \cmark \\
\rowcolor{autogategreen} deframer\_\autogate  & 82.31  & \textbf{77.29}  & \textbf{$-$17.65\%} & 13.77  & +1.47\%  & 100\% & 55.80\% & 7 & 0.00 & \cmark \\
\midrule
\rowcolor{basegray} mrisc\_base         & 729.80 & 664.45 & ---     & 184.92 & ---      & 84.00\%  & 80.40\% & 89 & 0.06 & \cmark \\
mrisc\_POET         & 726.60 & 661.10 & $-$0.50\%   & 185.38 & +0.25\%  & 84.00\%  & 80.40\% & 89 & 0.07 & \cmark \\
mrisc\_ROVER        & 727.90 & 662.49 & $-$0.29\%  & 184.44 & $-$0.26\% & 84.30\% & 80.50\% & 90 & 0.00 & \cmark \\
\rowcolor{autogategreen} mrisc\_\autogate     & 380.80 & \textbf{310.41} & \textbf{$-$53.28\%}  & 201.42 & +8.92\%   & 95.10\%  & 90.70\% & 104 & 0.06 & \cmark \\
\bottomrule
\end{tabular}
\end{table*}

\begin{figure*}[t]
\centering
\includegraphics[width=0.9\textwidth]{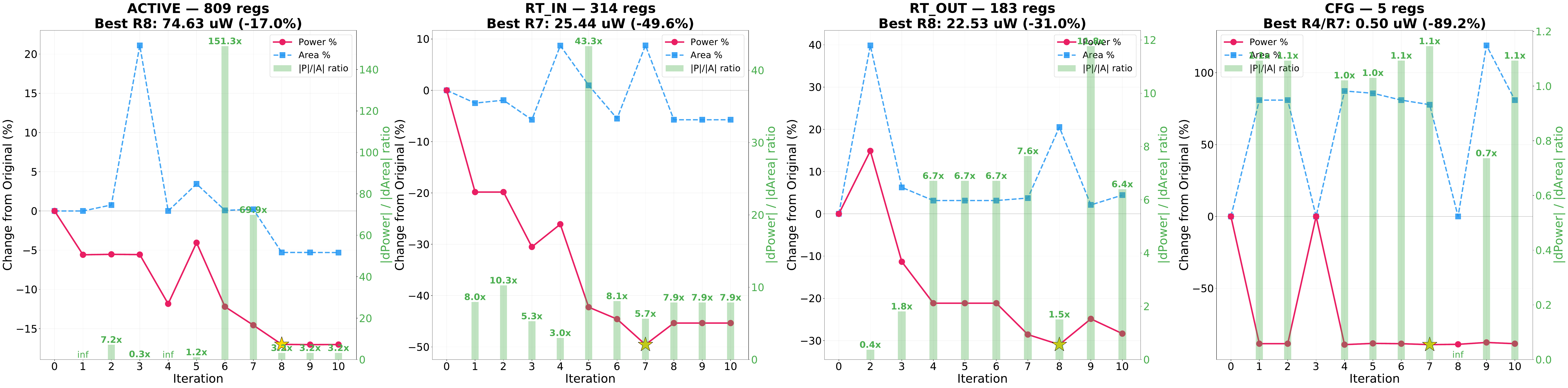}
\vspace{-10pt}
\caption{Power vs.\ area tradeoff across the four optimized partition M sub-modules. Each point represents a candidate rewrite; the selected variants (star) achieve the best power--area balance.}
\label{fig:cmac_power_area}
\vspace{-14pt}
\end{figure*}

\vspace{-5pt}
We evaluate \autogate on six small RTL benchmarks from CVDP~\cite{cvdp}, OpenCores~\cite{opencores}, and OpenTitan~\cite{opentitan}, two industrial-scale open-source designs from NVDLA~\cite{nvdla_hw} and BlackParrot~\cite{petrisko2020blackparrot}, and two highly optimized proprietary production designs.
Table~\ref{tab:benchmarks} summarizes the characteristics of the open-source designs.


We compare \autogate against two recent RTL power-optimization approaches, POET\cite{ping2026poet} and ROVER\cite{coward2024combining}. Since publicly available implementations are unavailable, we re-implement their approaches and evaluate them using the same synthesis, simulation, SAIF generation, PrimePower analysis, formal verification, and timing constraints as \autogate. All reported power and area changes are measured relative to the original baseline RTL design.

POET and ROVER have primarily been evaluated on small designs and do not provide a repository-scale optimization flow. For comparison on large designs, we apply POET and ROVER to the same modules and partitions selected by \autogate's hierarchy analysis. The resulting RTL modifications are then merged and evaluated under the same full-design workloads as \autogate.
Both \autogate and the POET baseline use Claude Opus 4.7 with the same token budget. The POET baseline was configured with a population size of $N=10$, maximum generations $G=10$, offspring $L=10$ per generation, and three repair attempts per candidate.


All designs were synthesized in a commercial 3\,nm-class technology library using the design-specified target clock period (or 1\,ns when unspecified), and use the most aggressive automatic clock-gating setting with minimum clock-gating bit-width = 4. Power was evaluated from gate-level SAIF traces and correctness was verified using commercial formal equivalence checking. These settings match the production synthesis flow used in industrial design practice. All reported savings arise solely from RTL rewrites that expose clock-gating opportunities beyond those already inferred by synthesis.



\subsection{Small Benchmark Results}



Table~\ref{tab:cvdp} reports results on small benchmarks ranging from single-file datapath blocks to multi-file processor RTL under diverse baseline clock-gating conditions. Since these designs do not provide testbenches, we leverage Claude Opus 4.7 to generate the corresponding synthetic testbenches and ensure that they achieve at least 95\% functional coverage with a commercial simulation tool.



\noindent\textbf{Key Observations.}
The largest savings occur in designs with limited clock-gating efficiency. For example, booth multiplier (booth\_mult), \autogate cuts dynamic power by 86.21\% by restructuring register assignments and raising clock-gating efficiency from 8.30\% to 87.00\%. Counter splitting and enabling refactoring lift clock-gating efficiency from 22.90\% to 82.80\% for counter32 and from 15.40\% to 81.50\% for divide. And Mini-risc(mrisc) achieves a 53.28\% reduction via multi-file coordination.


\subsection{Large Benchmark Results}

Table~\ref{tab:nvdla_combined} reports NVDLA partition-level results under two representative testbench workloads: \texttt{dc\_6x8x192} (Workload~1) and \texttt{dc\_35x22x54} (Workload~2). Optimization is performed under the Workload~1 testbench, while Workload~2 evaluates cross-workload transferability. BlackParrot results are evaluated under the Dhrystone testbench workload.

\paragraph{POET and ROVER results}

POET and ROVER remain close to the baseline on both NVDLA and BlackParrot, showing limited scalability to large hierarchical RTL designs. POET is constrained by long waveform traces and repository-scale codebases that exceed practical LLM context limits, while ROVER relies on predefined rewrite rules and lacks workload-aware analysis. Consequently, both achieve limited power improvements.

\paragraph{\autogate results}
\autogate overcomes these limitations through switching-aware FGCG analysis algorithm and targeted, hierarchical rewriting. On NVDLA, it reduces dynamic power across all partitions under both workloads, with the largest savings on Partition\_M (Workload~1, $-47.60\%$) and Partition\_A (Workload~2, $-33.93\%$). Partition\_C shows consistent reductions with 12.64\% area savings, while Partition\_P achieves smaller savings due to already high baseline efficiency (99.73\%). Improvements under Workload~2 demonstrate cross-workload transferability. On BlackParrot, \autogate achieves a 7.96\% dynamic power reduction with only 0.04\% area overhead. Additionally, clustering-guided merging of ICGs can reduce area while improving power.

\begin{table*}[t]
\centering
\caption{Combined large benchmark results. The upper portion reports NVDLA partition results under two workloads; the lower portion reports full-design BlackParrot results. Workload-dependent NVDLA columns are shown under each workload, while shared NVDLA columns are shown once.}
\vspace{-5pt}
\label{tab:nvdla_combined}
\renewcommand{\arraystretch}{0.6}
\footnotesize
\setlength{\tabcolsep}{1.3pt}
\begin{tabular}{l r r r r r r r r r r r r c c}
\toprule
\multicolumn{15}{@{}l}{\textbf{\textit{NVDLA partition results}}} \\
\midrule[0.08em]
& \multicolumn{4}{c}{\textbf{Workload 1: dc\_6x8x192}} & \multicolumn{4}{c}{\textbf{Workload 2: dc\_35x22x54}} \\
\cmidrule(lr){2-5} \cmidrule(lr){6-9}
\textbf{Design} & \textbf{\makecell{Total\\(uW)}} & \textbf{\makecell{Dyn\\(uW)}} & \textbf{Dyn $\Delta$} & \textbf{CG Eff} & \textbf{\makecell{Total\\(uW)}} & \textbf{\makecell{Dyn\\(uW)}} & \textbf{Dyn $\Delta$} & \textbf{CG Eff} & \textbf{\makecell{Area\\(um\textsuperscript{2})}} & \textbf{Area $\Delta$} & \textbf{CG Cov} & \textbf{ICGs} & \textbf{\makecell{WNS\\(ns)}} & \textbf{\makecell{Timing\\Pass}} \\
\midrule
\rowcolor{basegray} Part\_M\_base & 293.60 & 89.40 & --- & 96.92\% & 755.80 & 547.90 & --- & 69.41\% & 653.04 & --- & 99.89\% & 135 & 0.09 & \cmark \\
Part\_M\_POET & 293.60 & 89.40 & 0.00\% & 96.92\% & 755.80 & 547.90 & 0.00\% & 69.41\% & 653.04 & 0.00\% & 99.89\% & 135 & 0.09 & \cmark \\
Part\_M\_ROVER & 293.60 & 89.39 & $-$0.01\% & 96.92\% & 755.80 & 548.00 & +0.02\% & 69.41\% & 653.04 & 0.00\% & 99.89\% & 135 & 0.00 & \cmark \\
\rowcolor{autogategreen} Part\_M\_\autogate & 230.20 & \textbf{46.80} & \textbf{$-$47.60\%} & 97.83\% & 646.40 & \textbf{462.50} & \textbf{$-$15.60\%} & 68.89\% & 606.77 & $-$7.10\% & 99.87\% & 34 & 0.06 & \cmark \\
\midrule
\rowcolor{basegray} Part\_C\_base & 1187.00 & 414.80 & --- & 93.12\% & 1142.00 & 371.10 & --- & 93.93\% & 2432.80 & --- & 96.37\% & 497 & 0.00 & \cmark \\
Part\_C\_POET & 1188.00 & 415.40 & +0.10\% & 93.12\% & 1142.00 & 371.10 & 0.00\% & 93.93\% & 2433.41 & +0.03\% & 96.37\% & 497 & 0.00 & \cmark \\
Part\_C\_ROVER & 1188.00 & 414.70 & $-$0.02\% & 93.12\% & 1143.00 & 371.10 & 0.00\% & 93.93\% & 2433.50 & +0.03\% & 96.37\% & 497 & 0.00 & \cmark \\
\rowcolor{autogategreen} Part\_C\_\autogate & 1040.00 & \textbf{372.60} & \textbf{$-$10.17\%} & 93.02\% & 1003.00 & \textbf{336.00} & \textbf{$-$9.46\%} & 93.82\% & 2125.25 & $-$12.64\% & 95.73\% & 355 & 0.00 & \cmark \\
\midrule
\rowcolor{basegray} Part\_A\_base & 842.70 & 164.70 & --- & 99.24\% & 1378.00 & 700.40 & --- & 94.38\% & 2148.94 & --- & 99.83\% & 510 & 0.00 & \cmark \\
Part\_A\_POET & 842.70 & 164.70 & 0.00\% & 99.24\% & 1378.00 & 700.40 & 0.00\% & 94.38\% & 2148.08 & $-$0.04\% & 99.83\% & 510 & 0.00 & \cmark \\
Part\_A\_ROVER & 842.70 & 164.70 & 0.00\% & 99.24\% & 1378.00 & 700.40 & 0.00\% & 94.38\% & 2148.94 & 0.00\% & 99.83\% & 510 & 0.00 & \cmark \\
\rowcolor{autogategreen} Part\_A\_\autogate & 834.00 & \textbf{114.10} & \textbf{$-$30.72\%} & 99.73\% & 1252.30 & \textbf{462.75} & \textbf{$-$33.93\%} & 95.77\% & 2276.63 & +5.94\% & 99.83\% & 524 & 0.00 & \cmark \\
\midrule
\rowcolor{basegray} Part\_P\_base & 2695.00 & 43.52 & --- & 99.73\% & 2700.00 & 47.15 & --- & 99.73\% & 9198.16 & --- & 99.79\% & 1190 & 0.00 & \cmark \\
Part\_P\_POET & 2695.00 & 43.53 & +0.02\% & 99.73\% & 2700.00 & 47.18 & +0.10\% & 99.73\% & 9197.98 & $-$0.09\% & 99.79\% & 1190 & 0.00 & \cmark \\
Part\_P\_ROVER & 2697.00 & 43.54 & +0.05\% & 99.73\% & 2702.00 & 47.20 & +0.11\% & 99.73\% & 9205.38 & +0.08\% & 99.79\% & 1190 & 0.00 & \cmark \\
\rowcolor{autogategreen} Part\_P\_\autogate & 2695.00 & \textbf{41.12} & \textbf{$-$5.52\%} & 99.75\% & 2700.00 & \textbf{44.71} & \textbf{$-$5.17\%} & 99.74\% & 9194.29 & $-$0.04\% & 99.79\% & 1191 & 0.00 & \cmark \\
\midrule
\multicolumn{15}{@{}l}{\textbf{\textit{BlackParrot full-design results}}} \\
\midrule[0.08em]
\multicolumn{2}{l}{\textbf{Design}} & \textbf{\makecell{Total Pwr\\(uW)}} & \textbf{\makecell{Dyn Pwr\\(uW)}} & \textbf{Dyn $\Delta$} & \multicolumn{2}{r}{\textbf{\makecell{Area\\(um\textsuperscript{2})}}} & \textbf{Area $\Delta$} & \textbf{CG Cov} & \textbf{CG Eff} & \textbf{ICGs} & \multicolumn{2}{r}{\textbf{\makecell{WNS\\(ns)}}} & \multicolumn{2}{c}{\textbf{\makecell{Timing\\Pass}}} \\
\midrule
\rowcolor{basegray} \multicolumn{2}{l}{BlackParrot base} & 72,600 & 28,950 & --- & \multicolumn{2}{r}{129,863.65} & --- & 99.39\% & 99.25\% & 46,398 & \multicolumn{2}{r}{0.00} & \multicolumn{2}{c}{\cmark} \\
\multicolumn{2}{l}{BlackParrot POET} & 72,600 & 28,955 & +0.02\% & \multicolumn{2}{r}{129,876.60} & +0.01\% & 99.39\% & 99.25\% & 46,398 & \multicolumn{2}{r}{0.00} & \multicolumn{2}{c}{\cmark} \\
\multicolumn{2}{l}{BlackParrot ROVER} & 72,600 & 28,957 & +0.02\% & \multicolumn{2}{r}{129,830.19} & $-$0.03\% & 99.39\% & 99.25\% & 46,398 & \multicolumn{2}{r}{0.00} & \multicolumn{2}{c}{\cmark} \\
\rowcolor{autogategreen} \multicolumn{2}{l}{BlackParrot \autogate} & 70,000 & \textbf{26,645} & \textbf{$-$7.96\%} & \multicolumn{2}{r}{129,921.20} & +0.04\% & 99.52\% & 99.45\% & 44,291 & \multicolumn{2}{r}{0.00} & \multicolumn{2}{c}{\cmark} \\
\bottomrule
\end{tabular}

\end{table*}

\subsection{Module-Level Analysis}


Table~\ref{tab:partc_hierarchy} reports the per-module results for Partition\_C under the dc\_6x8x192 workload, with indentation indicating module hierarchy. Of the 43 RTL modules in Partition\_C, only 7 were selected for optimization based on power-bottleneck analysis.
Across these modules, \autogate achieves substantial power reductions with modest area overhead, while non-critical modules are retained at baseline.

Figure~\ref{fig:cmac_power_area} visualizes the power--area tradeoff across the optimized Partition M submodules across iterations.
Each point is a formally verified candidate, and the highlighted selections show how \autogate chooses variants that reduce dynamic power while constraining area growth.
\vspace{-5pt}
\begin{table}[h!]

\centering
\caption{Per-module power and area results for NVDLA Partition\_C.}

\label{tab:partc_hierarchy}
\renewcommand{\arraystretch}{0.1}
\scriptsize
\begin{tabularx}{0.9\columnwidth}{X  r @{\hspace{2em}} r}
\toprule
\textbf{Module} & \textbf{\makecell{Dynamic Power $\Delta$}} & \textbf{Area $\Delta$} \\
\midrule
Partition\_C
  & $-$10.17\% & $-$12.64\% \\
\hspace{1.5em}\textcolor{brown}{\textbf{$\vert$----}}\;CDMA\_dc
  & $-$6.19\%  & $+$0.05\%  \\
\hspace{1.5em}\textcolor{brown}{\textbf{$\vert$----}}\;CDMA\_wt
  & $-$5.65\%  & $-$1.81\%  \\
\hspace{3.0em}\textcolor{brown}{\textbf{$\vert$}}\hspace{0.8em}\textcolor{brown}{\textbf{$\vert$----}}\;WT\_8ATMM\_fifo
  & $-$9.87\%  & $-$0.33\%  \\
\hspace{1.5em}\textcolor{brown}{\textbf{$\vert$----}}\;CSC\_sg
  & $-$58.70\% & $+$1.01\%  \\
\hspace{3.0em}\textcolor{brown}{\textbf{$\vert$}}\hspace{0.8em}\textcolor{brown}{\textbf{$\vert$----}}\;SG\_dat\_fifo
  & $0.00\%$   & $-$9.96\%  \\
\hspace{1.5em}\textcolor{brown}{\textbf{$\vert$----}}\;CSC\_dl
  & $-$31.39\% & $-$0.59\%  \\
\hspace{1.5em}\textcolor{brown}{\textbf{$\vert$----}}\;CSC\_wl
  & $-$5.08\%  & $-$0.57\%  \\
\bottomrule
\end{tabularx}
\end{table}

\subsection{Ablation Study}

\begin{figure}[H]
\centering
\includegraphics[width=0.85\columnwidth]{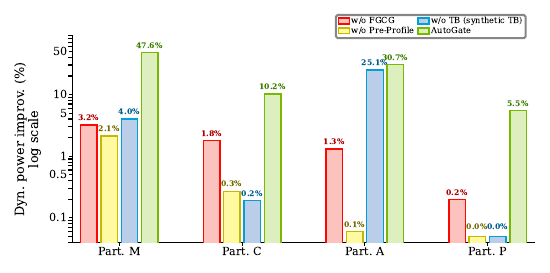}
\vspace{-1.5em}
\caption{Ablation study on NVDLA partitions in log scale.}
\label{fig:ablation}
\end{figure}

\textbf{Ablation I: AUTOGATE Components.}
To quantify the contribution of each major component, we perform an ablation study on the four NVDLA partitions under the dc\_6x8x192 workload. We evaluate three ablations: (1) \textbf{w/o FGCG}, where the agent receives no toggle-based clustering report; (2) \textbf{w/o Pre-Profile}, where the agent loses power bottleneck detection and hierarchy guidance; and (3) \textbf{w/o TB}, where synthetic per-module testbenches replace extracted workload-driven stimuli. Fig.~\ref{fig:ablation} reports the resulting dynamic-power improvements in log scale. The full \autogate achieves the best results.

\begin{figure}[H]
\centering
\includegraphics[width=0.85\columnwidth]{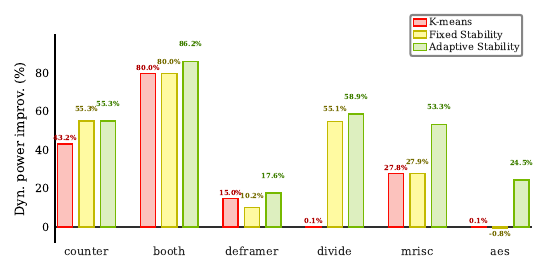}
\vspace{-1.5em}
\caption{Comparison of dynamic power reduction achieved by K-means, fixed-stability, and Adaptive Stability (ours) on small benchmarks.}
\label{fig:algo_power_comparison}
\end{figure}

\textbf{Ablation II: FGCG Clustering Algorithm.}
To isolate the impact of the FGCG clustering algorithm, we compare three clustering strategies while keeping the remainder of the \autogate flow unchanged. Fig.~\ref{fig:algo_power_comparison} compares the proposed adaptive stability clustering against K-means and fixed-threshold stability clustering~\cite{le2015stability}. Adaptive stability clustering consistently achieves the largest dynamic-power reduction, demonstrating the effectiveness of automatic threshold discovery.

\smallskip

\subsection{Proprietary Production Design Results}
To evaluate generalization beyond open-source benchmarks, we apply \autogate to two highly optimized proprietary production designs that are several times larger than NVDLA and BlackParrot. Table~\ref{tab:internal} reports dynamic-power and area changes across workloads. ``Single'' indicates that \autogate uses only Workload~1 for FGCG optimization, while ``Combined'' indicates that the optimization target includes both Workload~1 and Workload~10, representing two distinct activity patterns. For Design~A, single-workload optimization improves 9/13 workloads and achieves a 5.92\% average dynamic-power reduction, while combined optimization improves 12/13 workloads and achieves a 6.86\% reduction. For Design~B, the corresponding results improve from 6/13 workloads and 0.29\% average reduction to 12/13 workloads and 0.45\% average reduction. Area overhead remains below 1\% in all cases. These results indicate that optimizing over more diverse workloads can mitigate potential overfitting to a specific workload. A more detailed discussion is provided in Section~\ref{sec:limitations}.
\vspace{-10pt}
\begin{table}[H]
\centering
\caption{Dynamic power and area change on proprietary production designs across all workloads. }
\vspace{-5pt}

\label{tab:internal}
\scriptsize
\renewcommand{\arraystretch}{0.3 }
\setlength{\tabcolsep}{0.9pt}
\begin{tabular*}{\columnwidth}{@{\extracolsep{\fill}}lrrlrr@{}}
\toprule
\multicolumn{3}{c}{\textbf{\textit{Design~A}}} & \multicolumn{3}{c}{\textbf{\textit{Design~B}}} \\
\midrule
\textbf{Workload} & \textbf{Single} & \textbf{Combined} & \textbf{Workload} & \textbf{Single} & \textbf{Combined} \\
\midrule[0.08em]
Workload~1  & $-$3.72\%~\textcolor{green!60!black}{\(\downarrow\)}  & $-$2.83\%~\textcolor{green!60!black}{\(\downarrow\)}  & Workload~1  & $-$2.01\%~\textcolor{green!60!black}{\(\downarrow\)} & $-$0.36\%~\textcolor{green!60!black}{\(\downarrow\)} \\
Workload~2  & $-$4.39\%~\textcolor{green!60!black}{\(\downarrow\)}  & $-$3.63\%~\textcolor{green!60!black}{\(\downarrow\)}  & Workload~2  & $-$2.85\%~\textcolor{green!60!black}{\(\downarrow\)} & $-$0.97\%~\textcolor{green!60!black}{\(\downarrow\)} \\
Workload~3  & $-$9.29\%~\textcolor{green!60!black}{\(\downarrow\)}  & $-$8.67\%~\textcolor{green!60!black}{\(\downarrow\)}  & Workload~3  & +1.36\%~\textcolor{red}{\(\uparrow\)} & $-$0.16\%~\textcolor{green!60!black}{\(\downarrow\)} \\
Workload~4  & $-$4.50\%~\textcolor{green!60!black}{\(\downarrow\)}  & $-$3.68\%~\textcolor{green!60!black}{\(\downarrow\)}  & Workload~4  & +1.24\%~\textcolor{red}{\(\uparrow\)} & $-$0.17\%~\textcolor{green!60!black}{\(\downarrow\)} \\
Workload~5  & $-$9.49\%~\textcolor{green!60!black}{\(\downarrow\)}  & $-$8.81\%~\textcolor{green!60!black}{\(\downarrow\)}  & Workload~5  & $-$2.93\%~\textcolor{green!60!black}{\(\downarrow\)} & $-$0.26\%~\textcolor{green!60!black}{\(\downarrow\)} \\
Workload~6  & $-$15.16\%~\textcolor{green!60!black}{\(\downarrow\)} & $-$14.92\%~\textcolor{green!60!black}{\(\downarrow\)} & Workload~6  & $-$6.50\%~\textcolor{green!60!black}{\(\downarrow\)} & $-$0.47\%~\textcolor{green!60!black}{\(\downarrow\)} \\
Workload~7  & $-$16.49\%~\textcolor{green!60!black}{\(\downarrow\)} & $-$16.23\%~\textcolor{green!60!black}{\(\downarrow\)} & Workload~7  & $-$1.38\%~\textcolor{green!60!black}{\(\downarrow\)} & $-$1.09\%~\textcolor{green!60!black}{\(\downarrow\)} \\
Workload~8  & $-$17.95\%~\textcolor{green!60!black}{\(\downarrow\)} & $-$17.66\%~\textcolor{green!60!black}{\(\downarrow\)} & Workload~8  & +0.10\%~\textcolor{red}{\(\uparrow\)} & 0.00\% \\
Workload~9  & $-$12.43\%~\textcolor{green!60!black}{\(\downarrow\)} & $-$12.92\%~\textcolor{green!60!black}{\(\downarrow\)} & Workload~9  & +3.03\%~\textcolor{red}{\(\uparrow\)} & $-$0.12\%~\textcolor{green!60!black}{\(\downarrow\)} \\
Workload~10 & +1.24\%~\textcolor{red}{\(\uparrow\)}    & $-$2.35\%~\textcolor{green!60!black}{\(\downarrow\)}  & Workload~10 & +3.47\%~\textcolor{red}{\(\uparrow\)} & $-$0.25\%~\textcolor{green!60!black}{\(\downarrow\)} \\
Workload~11 & +1.78\%~\textcolor{red}{\(\uparrow\)}    & $-$2.04\%~\textcolor{green!60!black}{\(\downarrow\)}  & Workload~11 & +2.88\%~\textcolor{red}{\(\uparrow\)} & $-$0.84\%~\textcolor{green!60!black}{\(\downarrow\)} \\
Workload~12 & +2.66\%~\textcolor{red}{\(\uparrow\)}    & $-$1.60\%~\textcolor{green!60!black}{\(\downarrow\)}  & Workload~12 & +0.09\%~\textcolor{red}{\(\uparrow\)} & $-$0.29\%~\textcolor{green!60!black}{\(\downarrow\)} \\
Workload~13 & +10.73\%~\textcolor{red}{\(\uparrow\)}   & +6.14\%~\textcolor{red}{\(\uparrow\)}                 & Workload~13 & $-$0.24\%~\textcolor{green!60!black}{\(\downarrow\)} & $-$0.84\%~\textcolor{green!60!black}{\(\downarrow\)} \\
\midrule
\textbf{Average} & \textbf{$-$5.92\%} & \textbf{$-$6.86\%} & \textbf{Average} & \textbf{$-$0.29\%} & \textbf{$-$0.45\%} \\
\midrule
\textbf{Area $\Delta$} & \textbf{$-$0.38\%} & \textbf{+0.42\%} & \textbf{Area $\Delta$} & \textbf{+1.00\%} & \textbf{+0.02\%} \\
\bottomrule
\end{tabular*}

\renewcommand{\arraystretch}{1.0}
\end{table}

%% file: sections/conclusion.tex
\section{Limitations}
\label{sec:limitations}

AUTOGATE relies on workload-derived switching activity and may therefore overfit to the optimization workload. This workload dependence is especially evident on small designs, where the LLM can analyze the entire RTL design and testbench patterns, enabling workload-specific clock-gating optimizations that achieve up to 86.21\% power reduction. On large repository-scale designs, this effect is reduced because the increased design complexity and context size make it more difficult to specialize optimizations to a single workload, although the issue can still exist. As shown in Table~\ref{tab:internal}, optimizing for Workload~1 alone improves dynamic power under that workload, but can degrade results on Workloads~10--13. To mitigate this effect, multiple representative workloads can be incorporated during optimization. Although this may reduce savings on individual workloads, it improves overall power reduction and robustness across diverse workloads. An important direction for future work is the automatic identification of representative workloads that capture typical production use cases.

\section{Conclusion}
\label{sec:conclusion}

We presented \autogate, an agentic framework for automated RTL clock-gating optimization. \autogate demonstrates that ML--LLM co-designed RTL rewriting can achieve substantial dynamic-power reductions while scaling to industrial-scale RTL codebases, with marginal area overhead.